\documentclass[a4]{article}
\usepackage[a4paper]{geometry}
\usepackage{amsfonts,amssymb}

\usepackage{cite}
\usepackage[cmex10]{amsmath}
\usepackage{amssymb}
\usepackage[pdftex]{graphicx}
\usepackage{array}
\usepackage[tight,footnotesize]{subfigure}

\def\grad{{\overrightarrow\nabla}}

\def\intersect{{\;\cap\;}}

\usepackage{alltt}

\def\impos{{\;\vcenter{\hbox{\rule{5mm}{0.2mm}}} \vcenter{\hbox{\rule{1.5mm}{1.5mm}}} \;}}

\def\lrarrow{\leftrightarrow \kern-8pt \rightarrow}

\def\2{\frac{1}{2}}

\def\Nav{\langle N\rangle_{\overline T}}

\def\beq{\begin{eqnarray}}
\def\eeq{\end{eqnarray}}
\def\2{\frac{1}{2}}

\def\lrarrow{\leftrightarrow \kern-8pt \rightarrow}

\def\frightarrow{\rightarrow \kern-11pt /~~}
\def\reducesto{\simeq \kern -3pt >}

\begin{document}
\newcommand{\strust}[1]{\stackrel{\tau:#1}{\longrightarrow}}
\newcommand{\trust}[1]{\stackrel{#1}{{\rm\bf ~Trusts~}}}
\newcommand{\promise}[1]{\xrightarrow{#1}}
\newcommand{\revpromise}[1]{\xleftarrow{#1} }
\newcommand{\assoc}[1]{{\xrightharpoondown{#1}} }
\newcommand{\rassoc}[1]{{\xleftharpoondown{#1}} }
\newcommand{\imposition}[1]{\stackrel{#1}{\impos}}
\newcommand{\scopepromise}[2]{\xrightarrow[#2]{#1}}
\newcommand{\handshake}[1]{\xleftrightarrow{#1} \kern-8pt \xrightarrow{} }
\newcommand{\cpromise}[1]{\stackrel{#1}{\frightarrow}}
\newcommand{\policy}{\stackrel{P}{\equiv}}
\newcommand{\field}[1]{\mathbf{#1}}
\newcommand{\bundle}[1]{\stackrel{#1}{\Longrightarrow}}
\def\Server{{\rm Agent}_1}
\def\Client{{\rm Seed}}
\def\Proxy{{\rm Agent}_2}
\def\Dispatcher{{\rm Dispatcher}}

\title{A Promise Theory Perspective on\\The Role of Intent in Group Dynamics}
\date{\today}
\author{M. Burgess and R.I.M. Dunbar}
\maketitle

\renewcommand{\arraystretch}{1.4}

\begin{abstract}
  We present a simple argument using Promise Theory and dimensional
  analysis for the Dunbar scaling hierarchy, supported by recent data
  from group formation in Wikipedia editing. We show how the
  assumption of a common priority seeds group alignment until the costs
  associated with attending to the group outweigh the benefits in a
  detailed balance scenario. Subject to partial efficiency of
  implementing promised intentions, we can reproduce a series of
  compatible rates that balance growth with entropy.
\end{abstract}

\tableofcontents


\section{Introduction}

The dynamics of social groups in higher animals is deeply entwined
with cognition. For humans, social groups form around a hierarchy of
scales with very specific values \cite{dunbar6a}, resulting from the
way information flows between members of the groups
\cite{dunbar13a,dunbar14a}. In human groups, we one sees how the
frequency of individual contacts scales group sizes inversely with
frequency \cite{dunbar11a,dunbar12a} and this repetition amounts to a
sense of `trust' through increasing familiarity
\cite{dunbar10a,burgesstrust}.  A common base number for human groups
is about 5 individuals, which corresponds closely to the upper limit
of 4 on the number that can take part in a coherent conversation
\cite{dunbar15a,dunbar4a,dunbar8a,dunbar9a,dunbar16a}.  The limit on
conversational group size appears to be set directly by the capacity
to manage the mental states or viewpoints of other individuals
\cite{dunbar8a}. The proposal by Burgess to examine the role of trust
as a dynamical currency in social interactions between arbitrary
agents\cite{trustnotes} motivated an empirical study looking at group
phenomena on a large scale in Wikipedia editing; there the familiar
group patterns for humans were observed in an unusually large sample of data, and
could be viewed through the new theoretical lens of Promise Theory \cite{promisebook} for
calculating the sizes based on emergent scales \cite{burgessdunbar1}.

In this paper, we show how a theoretical model of contentious
autonomous agents, combined with straightforward physics of
dimensional analysis and some elementary statistical mechanics,
together offer an explanation for the main features of group dynamics.
In particular, we calculate the probability for reaching a certain
group size, based on the work expended in attenting to other agents.
The latter is clearly supported by (and acts as a proxy for) cognitive
capacity. We also show how this may be distorted in a predictable way
by the effects of non-human processes like `bots'. The results are in
good agreement with the data from over 200,000 individuals and bots.

Our model shows how groups come together in response to an initial
seed that attracts the attention of agents. The group accretes new
members until contention between them eventually drives the group
apart or the seed loses its interest value.  The predictions fit well
with the data from the Wikipedia study\cite{burgessdunbar1}, and give
credence to a new formalization of `trust' as a currency for
behavioural alignment rather than moral judgement.  Links to
neuroscience are also a natural place to seek explanations, for the
link between brain sizes and group sizes, and we point out a possible
connection to brain oscillation rates for different levels of attention.

\section{Statistical physics of human-machine agents}

Progress in theoretical social science has been slow compared to other
natural sciences.  Recently, attempts to model social phenomena, in
terms of variables that can be exposed as population characteristics,
has led to the nascent field of
Socio-Physics\cite{galam,sociophysics2}.  Socio-physics models argue
principally by analogy to known phenomena in physics (usually spin
models). However, a missing piece in these descriptions is the
underlying reasoning for both the variables and their likeness to
known problems in fundamental physics.  One imagines some universality
arguments at play, but without a deeper causal link, the likenesses
remain somewhat superficial in character. A second missing piece is
the vastly greater scope for semantics to play a role in behaviour on
a human scale. Discussions of elementary phenomena can afford to
suppress semantics and promote universality because there are few
degrees of freedom in play. Such isolation of information channels isn't
generally plausible in human systems unless one can argue in terms
of entropy.

Promise Theory, proposed by Burgess in the context of human-machine
systems\cite{burgessdsom2005}, was introduced to deal with such
criticisms in the context of technology. It takes issue with the
assumptions of logic as well as the suitability of Game
Theory\cite{myerson1}, Graph Theory\cite{berge1}, and Network
Science\cite{albert1}, but synthesizes all of these into a tool set
with fundamental principles based on compatibility with Information
Theory\cite{shannon1}. It was developed by Bergstra and Burgess over
the subsequent years\cite{promisebook}, and more recently has been
adopted as a model for socio-economic thinking.

The difficulty in formalizing human level concepts is that there is
often a tendency to fall back on moral philosophy or psychology to argue rather
than looking for an underlying causality on an agent or group 
level\cite{trustcoop,trusthandbook}. 
Promise Theory is a bottom up theoretical framework, embodying graph
theoretic notions as well as representations of semantic and dynamic
variability. It embodies founding principles for the autonomy of agents.
Taking a bottom up approach, it develops both algebraic formulations
and scaling principles.
Using Promise Theory, Burgess has proposed to use the familiar concept of trust as a
unifying instrument in order to connect familiar behavioural phenomena
(and terminology) with more formal analytical structures familiar from
physics. As we'll see, trust gives us a convenient dynamical potential onto which we
can graft the semantics of morality, through repeated processes like rituals
and beliefs. Thus, in this picture, we expect repetition to be a general principle
through which humans rehearse governance through social norms and graces.
These ritualistic promises provide surrogate `shared purposes' to rally people into collective behaviour.

\subsection{Intent and promises}

As outside observers, we can't always ask agents (people, animals, or
machinery) what drives them, or what they are thinking but we can try
to map their behaviours onto a model of intentions that match our own
thinking.  The intentional behaviour of an agent is rarely singular:
it could be a rich admixture of goals with varying priorities.  Each
agent will prioritize based on its own `algorithm'. Should a
particular seed promise pass a threshold attention level for group
formation, then this can manifest as an implicit bias, potentially
seeding an alignment with other agents.  In physics, we call this
spontaneous symmetry breaking; it's associated with phase transitions.

In Promise Theory, intentionality can be represented by stylized
`promises', which can be characterized by a `direction' of intent,
defined within a space of possible outcomes. We needn't expand on the
specific understanding of resources or processes that underly the
fulfillment of promises here. The key point is that this approach to
intent allows us to build an impartial algebraic representation for
each agent, without prejudicing the dynamics of changing intent.

An agent $A_1$ can promise something 
described by some intention (usually denoted $b$, for the body of the promise) 
in the form of an offer (denoted $+b$) to another agent $A_2$
\beq
A_1 \promise{+b} A_2.
\eeq
Assuming the agent $A_2$ accepts some amount of what is offered $b'$ (denoted $-b'$)
\beq
A_2 \promise{-b'} A_1,
\eeq
then a unidirectional flow of strength $b\intersect b'$ of intended outcomes binds the 
agents in a relationship.
This requires the attention of both parties, with donor and receptor promises, 
and involves time as an implicit resource.

Energy considerations alone are not a sufficient basis for determining
systemic behaviour: one also has to know the rules by which it gets
moved and exchanged between the dynamical variables of a system. Thus,
we need a story about the effective forces and representations for
interaction too. Our experiences in physics can help to shape the way
we represent these however.  

Promises are like the steady state solutions for motion in physical
systems.  In addition to promises, there can be transient events,
driven by changing boundary conditions or new information, such as when an agent throws a
ball to another to catch unannounced. These are denoted by a special
arrow resembling a fist:
\beq
A_1 \imposition{+b} A_2
\eeq
Impositions are transient (unplanned) interactions. They induce sudden demands or
costs onto the recipient,  they may be interpreted as the basis of contention, tending
to pull agents unpredictably off course.
In Promise Theory impositions
play a significant role in reducing trust\cite{accusations}. This 
dynamical picture is consistent with a
`work/energy' interpretation of trust alluded to in\cite{trustnotes}.
Promise Theory predicts that impositions tend to be ineffective, because they
are likely to be ill aligned with receptor promises. It also predicts that
impositions tend to reduce trustworthiness assessments and increase 
the attentiveness of a receiver.

Whether leading or following, agents that are imposed upon by their
episodic `neighbours' with transient demands tend to increase mistrust
or the attention level of the group.  This attracts agents with the
same concerns to align. Later they may drift away from the group when
the cost of attentive work exceeds a fractional threshold of group mistrust.
These are the elements we need to complete the basic model of group
formation.

\subsection{Promise patterns}

Simmel introduced the notion of triads in social
systems\cite{simmelsoc}. Another triad theory of sentiment relations
in social balance theory was proposed in
\cite{fritz1,fritz2,khanafiah2004social} as a rule of threes proposed for
the social sciences\cite{triads1}.  Triadic agent molecules have been
proposed many times as basic control units for social networks, and
have been implicated in group formation.  Promise Theory does not
recognize these speculative triadic network structures as such. Rather
it predicts a triangular co-dependence between agents arising from the
fundamental autonomy of agents. This means that agents fundamentally
make decisions from within, possibly tempered by conditions from
without, according to the law of assisted promise keeping\cite{promisebook}.
The pattern may be used to express the promise a message from sender to
receiver through a third party delivery agent, for example.

\begin{figure}[ht]
\begin{center}
\includegraphics[width=12cm]{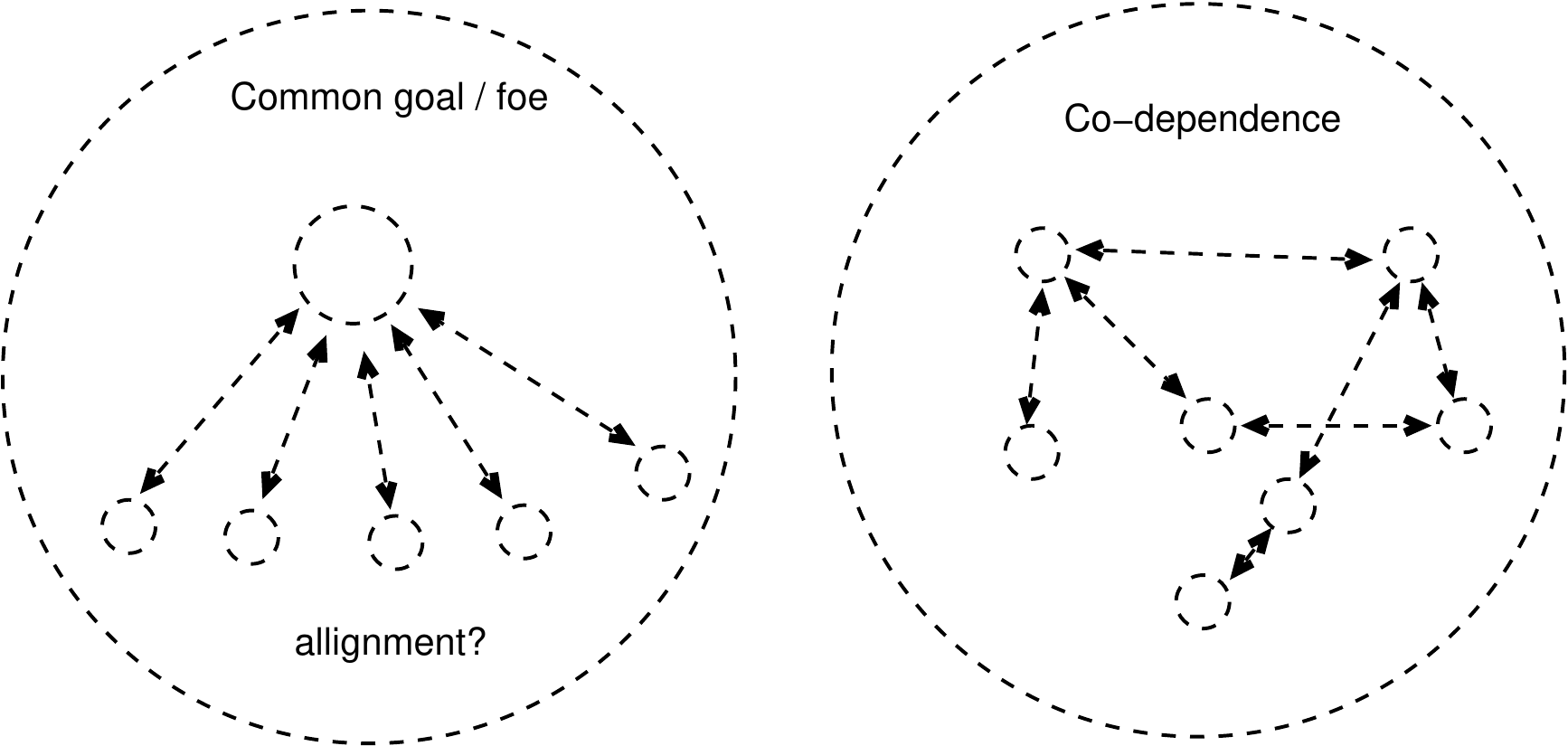}
\caption{\small Groups form either because agents come together
  independently attracted to contribute to a common cause (like
  fighting a common enemy or working on a common product), or they
  form emergent clusters by pairwise percolation of promise
  relationships. In our model, we assume the left hand picture of
  attraction in which `mistrust' of the central `seed' promise drives
  increased attention and potentially proximity as a secondary
  effect.\label{coop}}
\end{center}
\end{figure}

Figure \ref{coop} shows two ways in which groups could form.
The semantics of group formation are significant to the costs associated with them. A group that follows a single
leader or interacts one at a time is different from a group that tries
to maintain global coherence all at once and all the time. The latter
is very rigid and very expensive. For $N$ agents, the cost is of the order $N^2$.
In our discussion, we find the loose hierarchical association to be the cost that
provides the best agreement with data.

\begin{figure}[ht]
\begin{center}
\includegraphics[width=8cm]{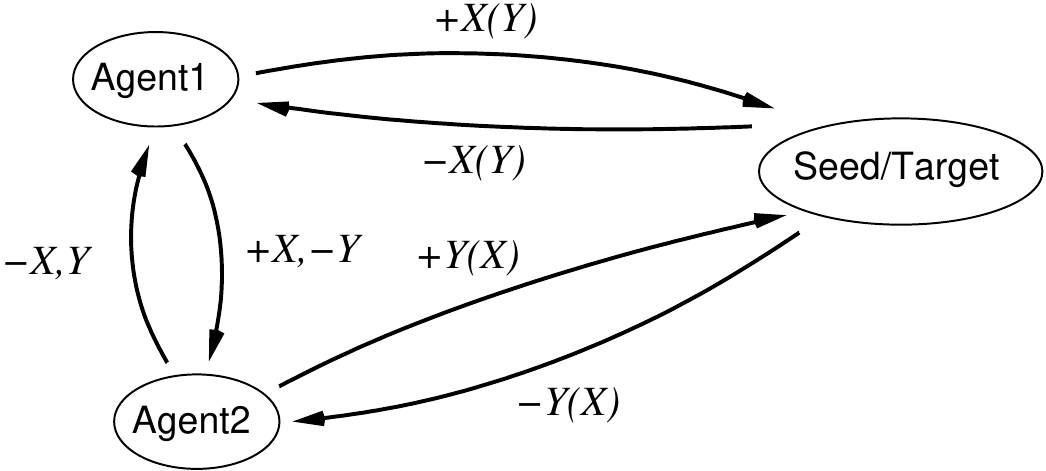}
\caption{\small A basic cooperation/calibration triangle in Promise Theory
allows two agents to work together on behalf of a third, or allows a third to act
as a seed effectively bringing them into alignment $X=Y$.
From Promise Theory, one would expect opportunistic dyadic structures 
$N=2$ for compositional or symbiotic specialization, 
with more important coordinated structures built from equilibrated/cross-checked
triads $N=3$.
\label{inter2}}
\end{center}
\end{figure}

Consider the primitive pattern involving three agents shown in figure
\ref{inter2}. The triangle of promises is the maximum
coordination for three agents. This is the configuration by which they
can maintain consistent information and claim to `agree' with one
another. It is called the Law of Conditional Assistance in Promise
Theory. It represents a configuration of voluntary cooperation respecting the
autonomy of the agents.  $\Server$ promises an intended
outcome $X$, based on the other agent's intent to supply $Y$' in the
most general sense.  The intended outcome $X$ could involve watching
over the group, performing some work on its behalf, etc. Essentially,
it requires paying attention to the promise and allocating time
resources.  $\Server$ also promise to make use of the promise $Y$
provided by $\Client$, which could simply be access to its personal
space, or the ability to perform some service for it.  We can use the
shorthand notation for the directed promises: \beq \left.
\begin{array}{c}
\pi_X: ~ \Server \promise{+X|Y} \Client\\
\pi_Y: ~ \Server \promise{-Y} \Client
\end{array}
\right\rbrace \equiv \Server \promise{+X(Y)} \Client.  
\eeq 
to represent the conditional promise of $X$ given $Y$, together with the promise to accept $Y$
if offered. In other words, `I will keep the promise of $X$ with the assistance of
another, who in turn helps me by supplying $Y$, written $+X|Y$, and I promise you that I am
accepting such help $-Y$'. Omitting the details\cite{promisebook}, the full collaboration now takes the form:
\beq
\Server &\promise{+X(Y)}& \Client \label{pull1}\nonumber\\  
\Server &\promise{-Y, +X}& \Proxy \label{pull4}\nonumber\\
\Proxy  &\promise{+Y,-X}& \Server \label{pull3}\nonumber\\
\Proxy &\promise{+Y(X)}& \Client \label{pull5}\nonumber\\
\Client &\promise{-X(Y)}& \Server \label{pull2}\nonumber\\
\Client &\promise{-Y(X)}& \Proxy\label{pull6}
\eeq
Notice the symmetries between $\pm$ in the promise collaboration of
equilibrium state, and between $X,Y$ indicating the complementarity
of the promises. 
The maximal cost of this configuration is close to the square of the number of agents.
Such a cost is unsustainable for large numbers.

For group cooperation, this level of cooperation is too expensive to
be sustained much beyond $n=3$, as the cost of predictable assurances
rises like $n^2$.  The level of cooperation we find represented in the
scaling of group formation data rather suggests only the association of
agents 1 and 2 through the proxy hub of the third. By extension, this is a classic
hierarchical model in which a single hub plays a coordination role in keeping
a group together---such as a head of department or group leader.
In other words, the Promise Theory, predicts that interactions we
might call `grooming' of the relationships are basically one to one
with a leader (scaling to one to $n-1$ in a group of size $n$),
as reflected in section \ref{accounting}).

When a seed is eliminated, or becomes overwhelmed by new priorities
arising from environmental pressures, contention between the seed
promise and impositions from the ambient environment rises,
destabilizing it as a priority. Thus, in the absence of a strong seed, there
is no effective promise to attract agents together and they drift
apart in response to the perturbations from competing intentions.

\section{Cost accounting for `grooming' work}

Our interpretation of trust is a pragmatic one. Ultimately, it's a
semantic abstraction of the `work of attention', or cognitive work, as
we'll show below.  Apart from minor semantic distinctions, such work
ranges over a variety of phenomena and scales from idle curiosity to
intense scrutiny and mistrust that are covered by the same attention
processes. We can call all of these forms of `kinetic mistrust'.

The differences are thus in the degrees of scrutiny and the
individual characterization of their importance of each agent. One
agent's causal interest might be another agent's response to
untrustworthy behaviour. We believe that this is consistent with
normal usage, but it allows us to formalize trust as a form of work
analogous to energy in two parts.

Since trust works as an attention accounting quantity,
it's driven by work done at different times, past and present.  As in
physics, potential energy is a summary of historically accumulated
work, expressed as a coarse snapshot of the slowly-varying history.
Conversely, kinetic energy is an immediate release of work, in
response to the tendencies of directionality expressed by the
potential.  Since potential is historically accumulated work, we need
{\em memory processes} to transmute learning into kinematics. Any
memory process will do, but agents that have brain matter are
obviously highly optimized for this and adds the sophistication for
dealing with memory on many levels from dynamics to semantics.

\subsection{Dimensional accounting}

Dimensional analysis is the way scales of measurement are defined in
natural science.  Classically, all measures can be reduced to
combinations of properties regarded as `innate' to physics, namely
mass, length, time, and a few others. The role of time is central to
group formation, principally because it is closely associated with
work.  The counting of any relationship with respect to time has to
follow a universal dimensional analysis. These relations were derived for
continuum processes for ballistic models.

In continuum language, a force $F$ applied over a path length $dx$ in
some parameter space is equivalent to a directional impulse $dp$. If one assumes a process velocity
$\vec v=\vec{dx}/dt$, where $\vec{dx}$ represents the direction of an intention in the space of outcomes,
 this settles the accounting of the quantities with respect to time. The usual
`Newtonian' conventions follow from the observation that a change in potential energy (defining a force) has the same
dimensions as a change in kinetic energy:
\beq
dV = \grad V\cdot\vec dx = \vec F\cdot \vec{dx} &=& \vec F\cdot \vec{v}\,dt\nonumber\\
&=& \frac{\vec{dp}}{dt}\cdot \vec v\;dt\nonumber\\
&=& \vec v\cdot \vec{dp}\nonumber\\
&=& m\vec v\cdot \vec{dv}\nonumber\\
&=& \2m d(\vec v\cdot \vec v)\nonumber\\
&=& d\left(\2 mv^2\right)\nonumber\\
&=& d\overline T.  \label{newton}
\eeq 
Thus, the relationship between the quantities we count as force and energy are constrained mainly by dimension
and rate, and energy accounting.

In our dynamics of potential alignments and kinetic 
attention processes, the kinetics are more like Shannon information sampling \cite{shannon1} than linear motion,
but the dimensions have to be the same. The rate of conversion of accumulated work from the keeping of promises
is thus found dimensionally by comparing
\beq
V \sim \2 mv^2
\eeq
up to dimensionless factors, where the 
attention rate or `velocity' $v$ whose dimensions are arbitrary except for the role of time.
The potential amounts to a reliability for promise keeping. One might even call it a kind of `goodwill' $V$
in human terms.
The work of a single agent, interacting in a group of size $n$, would be expected to scale as
\beq
W\text{(agent)} = \frac{c_1+c_2(n-1)}{c_0 n_\beta}
\eeq
where $c_0$, $c_1$, and $c_2$ are constants.
At low utilization, we can expect the availability or channel capacity to be approximately proportional to
the number of agents interacting. Once contentions sets in, this effective number slows down
as agents begin to leave a group to an average---which is the value at which contention is maximal.

When $\beta E = 0$, the probability has to be 1, so for $n=1$ (self), all the share is
in one agent's hands. So $c_1=0$. Now we have a single scale $C\equiv c_2/c_0$ representing
the level of shared of contention between agents. To determine this, we use the promise
seed configuration again below. Note that, at maximum entropy, this is evenly distributed
without particular favour to any agent. So, based on these dimensional arguments, 
we expect the limit of maximum entropy for large $N$ to take the form:
\beq
P(\beta) \sim \exp\left(-\frac{C(n-1)}{n_\beta}\right),
\eeq
where $n_\beta$ is some scale that characterizes the intra-group contention, Small $C$ 
implies tolerance of contention, or loose coupling and thus larger group sizes,
while large $C$ implies some kind of territorial overlap that leads to altercation.

\subsection{Work afforded by a limited capacity}\label{accounting}

We can make this more formal as follows. Suppose each agent has a cognitive processing work capacity $W_\text{max}$
for the process of group interactions that it shares with other tasks too. How the capacity is sliced
is a detail that we don't need to address here. 
If we think in terms of the `power output' or work of the agents, in kinetic terms,
relating to the promise of sharing the group resources. At max entropy (large $N$ and large ensembles), 
the probable work fraction $P(W)$ for distribution would take the form of a Boltzmann distribution over the relative
costs\cite{reif1,treatise1}:
\beq
P &\sim& e^{-\beta E}\\
\text{where} ~~~ \beta E &\mapsto& \frac{W(\text{agent})}{\text{Total capacity}},
\eeq
where the availability is the finite budget for shared resource channel capacity. Moreover, from Shannon\cite{shannon1}, 
we know that the channel capacity is a dimensionless representation 
of the power:
\beq
C = B \log \left( 1 + \frac{W(\text{agent})}{\text{Cost of contention}}\right)
\eeq
where $B$ is the maximum bandwidth for throughput. 

With these points in mind, and assuming that interactions between group members are not `all at once', 
but interleaved principally one at a time,
the accumulated work should be proportional to the group remainder size:
\beq
W_n \le \frac{W_\text{max}}{n},
\eeq
The bulk of this work is assumed to be the handling of impositions by unexpected group members
to reverse efforts and otherwise interfere with the agent concerned, preventing or smoothing over such incidents.
The agent may have other things to deal with in addition to `grooming' or placating contentious others, so this work
allocation might not be 100\% efficient.
So we can take the cognitive capacity as a share for work:
\beq
(n-1)W_n = \2m v^2,
\eeq
for some rate $v$. Now, we arrange to measure these quantities in units such that we can
compare dimensionless ratios. In dimensionless form, we can write
\beq
(n-1)\frac{W_n}{W_\text{max}} = \2 \frac{m}{m_\text{min}} \left(\frac{v}{v_\text{max}}\right)^2,
\eeq
The effective mass of the interaction (which plays the role of the 
cost of agent ``involvement'' with others) presumably has a minimum scale rather than a maximum,
though this doesn't matter since we eliminate this by changing variables.
None of these work rates are measurable in this study, so we need to relate them to something
with dimensions of $n$. We can make the identification
\beq
\frac{W_n}{W_\text{max}} \frac{m_\text{min}}{m} \equiv \frac{\beta}{\Nav}, \label{xx}
\eeq
which has the form
\beq
\frac{\text{Fractional work effort}}{\text{Fractional cost of involvement}}\times \text{efficiency},
\eeq
where we use the constant $\beta \le 1$ as an efficiency.
This is motivated by the identification of $\Nav$ as the scale for group size with
maximal contention cost.
From (\ref{xx}) we interpret the Dunbar group size as being based on:
\beq
\Nav = \text{cost as a fraction of work budget} \times \text{efficiency}.
\eeq
In other words, $\Nav$ may be called the group contention cost, measured in cognitive work units.
The actual values for $\Nav$ can't be derived without a specific implementation model, but
we expect this is an innate internal capacity of each kind of agent, as originally
proposed by Dunbar. In that case, this dimensional identification, based on the assumption of 
linearly shared work together with the assumption of maximal entropy yields results compatible
with the social brain hypothesis. The fact that we rediscover this in the case of Wikipedia histories\cite{burgessdunbar1}
is evidence that the editing is a principally human to human interaction, albeit with cyborg influences.
In the Wikipedia results, $\beta = 1$ gives the appropriate fit.
In Dunbar's human groups. $\beta=0.75$ is a closer estimate of the promise efficiency.

Agents come together around a particular seed when their
prioritization of the seed promise becomes the dominant force in their
behaviour. For example, the appearance of a predator activates a
behaviour for a herd; the appearance of a new Wiki page on a subject
close to one's heart activates a desire to contribute.  In the absence
of an attraction, there are enough alternative attractions to pull
animals away, leading to an exponential decay of this heightened
priority.

\subsection{Probability of group size $n$}

\begin{figure}[ht]
\begin{center}
\includegraphics[width=12cm]{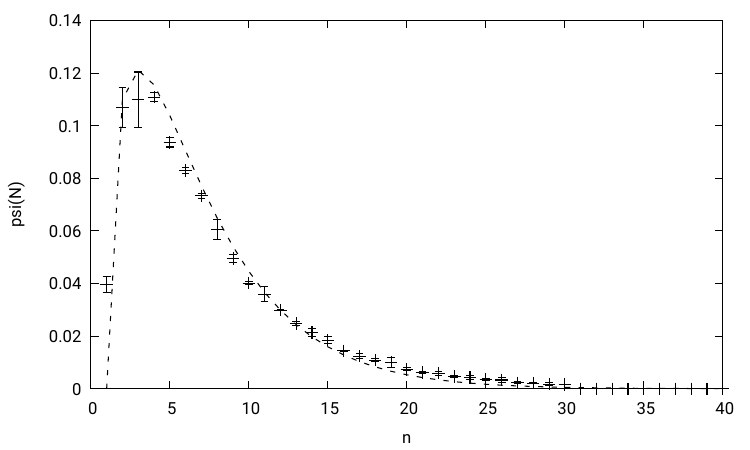}
\caption{\small Curve fit of data using the formula in equation \ref{formula}. The crosses
approximate error uncertainty. The model fit is expected to be worst for small $n$ due to 
integer effects.\label{datafit}}
\end{center}
\end{figure}

We can now extend this dimensional argument to predict the dimensionless frequency (or probability) of 
finding a group of size $n$, which we denote by $\psi(n)$.
The graph in figure \ref{datafit} fits very closely a simple formula which we can motivate from the theory:
\beq
\psi(\nu) = \frac{4}{\sqrt\pi}\; \frac{\nu^{\2}\; e^{-\nu}} {\langle N\rangle_{\overline T}}, 
~~~~~~\nu = \frac{2\beta(n-1)} {\langle N\rangle_{\overline T}} ~,~ (n > 1),\label{formula}
\eeq
where $\beta$ corresponds to a dimensionless (probabilistic) rate of promise keeping for the seed promise,
i.e. $\beta$ is the fraction of promises kept reliably, since reducing $\beta$ has the same effect as
reducing the group contention size limit higher (less tolerance of contention). Another way to think of $\beta$
is therefore as an metaphorical `temperature' complement for agent entropy. As contention increases, the
maximum occurs at smaller groups and that is equivalent to less effective promise keeping to interact with
the seed agent.
The result of this fit is shown in figure \ref{datafit}.

\subsection{The scaling of group hierarchy}

The precise fit of the formula \ref{formula} is subject to some tuning (see figure\ref{datafit} especially for small $n$).
The relationship between the maximum frequency and maximal contention scales is determined, however,
by the rate equation for detailed balance that leads to (\ref{formula}).
The value of $n$, which maximizes kinetic mistrust, is called $\langle N\rangle_{\overline T}$, while
the value of $n$ leading to the maximum value of $\psi(n)$, determined by $\frac{d\psi(N)}{dn}=0$ is:
\beq
n_i^\text{max} = 1 + \frac{\langle N\rangle_{\overline T}}{4\beta}.
\eeq
Notice how the expected group size is still always less than the maximal contention size.
This is interesting, as it suggests that (statistically) agents tend to prioritize working
more intimately with smaller groups. This could be a sign that there is an additional contention
cost associated with switching between on going relationships, as there is in computing called {\em context switching}.

We can examine some values for these maxima relationships 
to illustrate the fit with the layer model in Dunbar\cite{dunbar3}
and the specific data for Wikipedia\cite{burgessdunbar1}. The column for $\beta = 1$ reproduces the results from
the Wikipedia data in \cite{burgessdunbar1}. 
Removing all bot interactions arbitrarily alters $\Nav$ slightly to give an effective value of $\beta=0.93$.
The column with lower efficiency $\beta=0.875$ generates the usual stylized
Dunbar sequence quite accurately:
\begin{center}
\begin{tabular}{c|c|c|c}
Mode & Wiki & No Bots & Dunbar Approx \\
\hline
$n_i^\text{max}$ & $\langle N\rangle_{\overline T}~ (\beta=1)$& $\langle N\rangle_{\overline T}~ (\beta=0.93)$ & $\langle N\rangle_{\overline T}~ (\beta=0.875)$ \\
\hline
3 & 8 & &\\
5 & & 14.9 & 14\\
8 & 28 & &\\
14 & & & 45.5\\
14.9 & & 52 &\\
28 & 108 &&\\
45.5 & && 156\\
52 & & 188 &\\
108 & 428 &&\\
156 & & & 542\\
188 & & 697 &\\
428 & 1708 &&\\
542 && &1892\\
\end{tabular}
\end{center}

\begin{figure}[ht]
\begin{center}
\includegraphics[width=15cm]{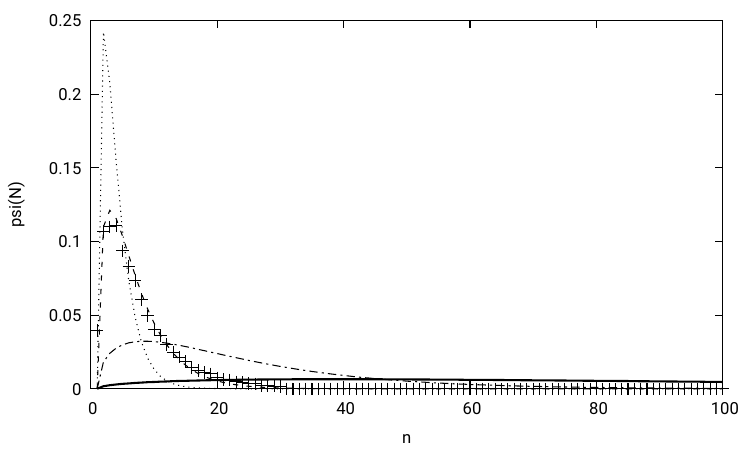}
\caption{\small The group equilibrium law plotted for $\langle N\rangle_{\overline T} = 4,8,30,150$
illustrating the flattening of group probability curves with increasing number. The amplitude
gives an approximate magnitude for the attention power rate required to maintain each level.\label{dunbar}}
\end{center}
\end{figure}

Reading down each column, we see the mode frequency limited by the next scale up in the two right hand columns.
We note that the apparent self-similar scaling fraction of group sizes depends on $\beta$
for its precise value. It's therefore unrelated to presence of triadic relationships in the promise graph
of agents, since the relevant promise graph is purely hub centric during group formation.
Some of the curves for these values are plotted in figure \ref{dunbar}.

\section{Human attention and neural processes}

We have shown that, if group size is moderated by contention, or grooming work to overcome it,
then achievable group size depends on the intrinsic timescales over which agents
can deal with the contention. This gets eliminated as a variable in 
probability distribution, but its remnants are found in dimensionless $\beta$. 
When the rate of seed-related promise keeping falls and $\beta$ falls in value, 
the group can only either sustain itself over a longer
timescale (through more invested work) or with fewer numbers.  If the work has
a maximum capacity, then the available fraction is spread more thinly
and less contentiously over larger $\langle N\rangle_{\overline T}$.

We can add a brief speculation to this prediction.  If the group
hierarchy is associated with cortical contention in humans (and indeed
other machinery), we should probably ask what are the dominant neural
processes at each level of the hierarchy? One possibility could
be that the group sizes correspond to different level of brain activity.
A proxy for these dynamics is perhaps `brainwave' oscillation modes\cite{buz0}
for the transport of information between cortical regions.

Frequency is associated with power\cite{motokawa}, so it's interesting to 
compare the hierarchy of group sizes to the power associated with levels of
attention or brain concentration. Buzs\'aki writes \cite{buz1}:
``The power density of local electrical field potential is inversely
proportional to frequency in the mammalian cortex.  This 1/f power
relationship implies that perturbations occurring at slow frequencies
can cause a cascade of energy dissipation at higher frequencies and
that widespread slow oscillations modulate faster local event.'' Thus
the idling work required for attentiveness in a typical group size
might be expected to follow the same kind of power
requirement.

Once again, on dimensional grounds $\Nav$ can only appear in this
relationship multiplied by an effective time conversion scale $\Delta\tau$
for the `latency', and the product of this with frequency  $f\times\langle N\rangle$ represents
an average throughput of information up to some intrinsic timescale $\Delta\tau$. So in relative units:

\begin{center}
\begin{tabular}{c|c|c|c}
Attention & Brain wave (Hz) $f$ & Dunbar $\langle N\rangle$ level & $f\times\langle N\rangle$\\
\hline
light attention& $\alpha$  5-15 (5)   & 150  & 750 \\
middle attention &$\beta$  12-30 (25)   & 30 & 750\\
 concentrated & $\gamma$-fast 32-200 (150)  & 5& 750\\
\end{tabular}
\end{center}
The product of the columns is approximately of constant order, 
suggesting that the average effort is indeed in inverse proportion to the
group size. This is numerically
interesting, if not exactly proof of a connection.

\section{Remarks}

What began as a pragmatic model of trust as attention in Promise
Theory has led us to a plausible explanation for the hierarchy of
social group sizes discovered by Dunbar. In this work, we bring
together these two narratives to offer a tantalizing perspective on
each.

The model makes a bold assumption, supported by the scaling, namely
that groups in a social brain hierarchy form around a seed of intent,
which acts to capture the attention of agents through associated
kinetic process.  There is a de facto attractive `force' that promotes
group accretion on a small scale, and later fades away to become asymptotically
free as groups disband.

Our summary relationship is based on continuum process algebra (usual
for large $N$), but we know that social groups are about individuals
(small $N$).  This is where the separation of scales in Promise Theory
is helpful. It is not the scaling of network that predicts these
results, but the underpinning process of assessment of social ties
that we call `attention' (and effectively `trust'). 
Thus a predictability of group behaviour requires the smooth
exchange of experiences over a large enough timescale to distill
stable patterns\footnote{At the time of writing, there is a
  revisitation of author Isaac Asimov's fictional Foundation Trilogy
  concerning the large scale prediction of human affairs, which he
  called `psychohistory'.  Although one can wonder whether Asimov
  would have been equipped to understand this, this kind of prediction
  over massive data sets and averages is the only level at which human
  behaviour is likely to be predictable.  Individual actions appear
  disconnected and noisy on a small scale.  His stories about robots
  imagined artificial brains that were ruled by potentials (Asimov was
  a biochemist) rather than digital logic, more like the cybernetics
  of Wiener and others.}. This separation of dynamical process scales is 
implicitly the result of evolutionary biology. Today, researchers in 
Artificial Intelligence dare to solve it with alternative models and computers.

What is the all important seed promise? By definition, it promises the
role of a prioritized behaviour that's shared by the individuals in a
group.  In the case of Wiki editing, it's clearly the promise of the
platform to enable satisfactory publishing of information---the
creative commons, with its attendant benefits.  For animals in a pack
or herd, it might be the promise of a defensive posture when a
predator is nearby, or the co-location of some tidbit, that drives them
to attend to one another's relative positions and cluster. They would
then drift apart again once the seed were gone\cite{jackdaw}.  For a religious group
or company, it could be a charismatic leader\cite{dunbar4}, which
also aligns with work on the origin and semantics of
authority\cite{burgessauthority1}.  Alternatively, it could be a more
abstract health benefit acquired as an evolutionary adaptation over
very long times, such as when a change in the weather or other
environmental conditions triggers group changes, as in slime mould
dissociation for instance---or merely the opportunistic sharing of a
transient resource\cite{ostrom1}.  The semantics of a seed of intent
might change frequently to reflect changing group dynamics, even as
the underlying dynamics remains a universal function of physiology.

Agents offer their attention to group processes variably in order to
invoke a simple optimization for beneficial reasons. They have a
finite budget for attention, which is governed by their work capacity.
In a future in which humans bond with artificial enhancements as
`cyborgs', Artificial Intelligence may alter some aspects of this.
This could, in turn, pose a different spectrum of threat to human
character that needs exactly the kind of cognitive capacity predicted
in the Dunbar hierarchy to deal with effectively.

\bigskip

Acknowledgement: MB is grateful to Gy\"orgy Busz\'aki for discussions about the neuroscience.
This work was supported by the NLnet Foundation.

\bibliographystyle{unsrt}
\bibliography{bib,spacetime}

\end{document}